\documentclass[aps,preprint]{revtex4}
\usepackage{graphicx}

\begin{document}


\title{Quantum impurity in a magnetic environment}

\author{Subir Sachdev}
\affiliation{Department of Physics, Yale University, P.O. Box
208120, New Haven CT 06520-8120 }

\date{\today}

\begin{abstract}
A unified perspective is given on a number of different problems
involving the coupling of a localized quantum spin degree of
freedom to the low energy excitations of an antiferromagnet, a
spin glass, or a Kondo insulator. The problems are related to
those in the class often referred to as ``Bose Kondo''.
\end{abstract}

\maketitle

\section{Introduction}
\label{intro}

The Kondo problem has played a central role in the development of
the theory of correlated electron systems. At its simplest it
consists of a single quantum spin, $\hat{S}_{\alpha}$
($\alpha=x,y,z$), interacting with the {\em fermionic\/}
excitations in a metallic environment. In a modern perspective,
many aspects of the Kondo problem can be understood in the
framework of boundary conformal field theory: the fermionic
excitations in the environment are represented by a $1+1$
dimensional, free, conformal field theory with central charge
$c=1$, and this interacts with the quantum spin degree of freedom
located at spatial co-ordinate $x=0$.

More recently, attention has focused on a new type quantum
impurity problem. Here, we again consider a quantum spin
$\hat{S}_{\alpha}$, but it now interacts with {\em bosonic\/}
excitations in the environment. Such models become appropriate
when the environment is in the vicinity of a magnetic ordering
transition, and there are low energy spin excitations in the bulk;
the latter may be viewed as excitonic particle-hole bound states
of a metal/insulator/superconductor which peel off below the
continuum of a pair of fermionic particles or holes.

We begin by describing the simplest `Bose Kondo' problem, and
postpone a discussion of specific physical motivations till later,
when we consider more realistic models. The simplest
model\cite{sy} has the Hamiltonian
\begin{equation}
\mathcal{H}_1 = -\lambda \phi_{\alpha} \hat{S}_{\alpha} \label{h1}
\end{equation}
where $\lambda$ is a coupling constant, and the $\hat{S}_{\alpha}$
obey the usual relations of a Heisenberg spin with angular
momentum $S$,
\begin{equation}
[ \hat{S}_{\alpha} , \hat{S}_{\beta} ] = i \epsilon_{\alpha \beta
\gamma} \hat{S}_{\gamma}~~~~;~~~~ \hat{S}_{\alpha}
\hat{S}_{\alpha} = S (S+1). \label{spin}
\end{equation}
The Bose field $\phi_{\alpha}$ has Gaussian correlations in the
absence of its coupling to $\hat{S}_{\alpha}$, with the two-point
correlation obeying
\begin{equation}
\langle \phi_{\alpha} (\tau) \phi_{\alpha} (0) \rangle_{\lambda=0}
\sim \frac{1}{|\tau|^{\mu}}, \label{corr}
\end{equation}
for large $|\tau|$ with $\mu > 0$, where $\tau$ is imaginary time.

It is important to distinguish the above Bose Kondo problem, from
the `spin boson' problem which had been the focus of much earlier
attention \cite{leggett}. The latter deals with a two-level system
coupled to a bath of harmonic oscillators. Upon interpreting the
two-level system as a spin, the splitting between the energy
levels behaves like a magnetic field on the spin. In this
situation, the spin-inversion symmetry, $\hat{S}_{\alpha}
\rightarrow -\hat{S}_{\alpha}$ for any 2 of the 3 $\alpha$ values,
is explicitly broken by the Hamiltonian. In contrast, in the Bose
Kondo problem of interest here, this spin inversion symmetry is
preserved (when combined with the transformation $\phi_{\alpha}
\rightarrow - \phi_{\alpha}$).

Despite the simple form of the Hamiltonian $\mathcal{H}_1$ and of
the correlator (\ref{corr}), the spin commutation relations
(\ref{spin}) make this a problem of some complexity which cannot
be solved exactly. This is also evident from its path integral
formulation, in which we integrate over $\phi_{\alpha} (\tau)$ and
over a unit length field $n_{\alpha} (\tau)$, where
$\hat{S}_{\alpha} = S n_{\alpha}$:
\begin{eqnarray}
&&\mathcal{Z}_1 = \int \mathcal{D} \phi_{\alpha} ( \tau) \mathcal{
D} n_{\alpha} (\tau ) \delta \left( n_{\alpha}^2 - 1 \right)
\exp\left(- \mathcal{S}_{\text{b}}
[\phi_{\alpha}] - \mathcal{S}_{\text{imp}} \right) \nonumber \\
&& \mathcal{S}_{\text{imp}} =  \int d \tau \bigg[ i S A_{\alpha}
(n) \frac{d n_{\alpha}(\tau)}{d \tau}  - \lambda S \phi_{\alpha}
(\tau) n_{\alpha} (\tau) \bigg] \nonumber \\
&& \mathcal{S}_{\text{b}}[\phi_{\alpha}] = \frac{1}{2}\int d \tau
d\tau^{\prime} \phi_{\alpha} (\tau) Q^{-1} (\tau - \tau^{\prime})
\phi_{\alpha} (\tau^{\prime}) . \label{simp}
\end{eqnarray}
In this formulation, all the non-linearities are in the first
Berry phase term, which involves the vector potential of a unit
Dirac monopole at the origin of spin space obeying
\begin{equation}
\epsilon_{\alpha\beta\gamma} \frac{\partial A_{\gamma}
(n)}{\partial n_{\beta}} = n_{\alpha}. \label{diracmono}
\end{equation}
Also, $Q(\tau)$ is the two-point $\phi_{\alpha}$ correlator at
$\lambda=0$, and (\ref{corr}) implies that its Fourier transform,
$Q(i \omega)$, has the spectral density $ \mbox{Im} Q(\omega) \sim
\mbox{sgn}(\omega) |\omega|^{\mu-1}$ at small frequencies.

The problem (\ref{simp}) appeared in Ref.~\onlinecite{sy} in the
context of a mean-field theory of a quantum Heisenberg spin glass.
It was solved here by generalizing the SU(2) symmetry to SU($N$),
and taking the large $N$ limit. In this limit, the spin
correlations obey
\begin{equation}
\langle \hat{S}_{\alpha} (\tau) \hat{S}_{\alpha} (0) \rangle \sim
\frac{1}{|\tau|^{2-\mu}} \label{e1}
\end{equation}
for large $\tau$ and $0 < \mu < 2$, while for $\mu \geq 2$ and
small $\lambda$ there is a broken spin rotation
symmetry\cite{anirvan} and the $\hat{S}_{\alpha}$ two-point
correlator reaches a non-zero value at large $|\tau|$. The
exponent in (\ref{e1}) was also obtained using a one-loop
renormalization group analysis \cite{anirvan,qmsi}, and was
subsequently shown \cite{sbv,gps} to hold to all orders in an
expansion in $2-\mu$. The result (\ref{e1}) has also been found to
hold in certain quantum impurity models in which the spin
$\hat{S}_{\alpha}$ is coupled simultaneously to bosonic and
fermionic excitations in its environment
\cite{anirvan,qmsi,si,demler}, including cases with spin
anisotropy.

It is important to note that (\ref{e1}) relies crucially on the
presence of the Berry phase in (\ref{simp}). In the absence of
this term, we can integrate over the Gaussian $\phi_{\alpha}$
modes, and then (\ref{simp}) becomes equivalent to a {\em
classical} ferromagnetic spin chain at finite `temperature', with
exchange interactions which decay as $1/|\tau|^{\mu}$. The
properties of this classical model\cite{kosterlitz} are very
different and have an interesting `dual' structure. Now, the
ferromagnetic phase with broken spin rotation symmetry appears
only for $\mu < 2$ and low `temperatures' (large $\lambda$)---in
contrast, with the Berry phase term, as we noted above, spin
rotation invariance was broken for $\mu \geq 2$ and small
$\lambda$. Furthermore, in the classical model without the Berry
phase, the paramagnetic phase with preserved spin rotation
symmetry (present for all `temperatures' for $\mu \geq 2$ and in
the high `temperature' (small $\lambda$) phase for $\mu < 2$) has
its two-point $\hat{S}_{\alpha}$ correlator decaying as
$1/|\tau|^{\mu}$---in contrast, with the Berry phase we found a
rotationally invariant phase for $\mu <2$ and with the correlator
(\ref{e1}).

Intriguing and interesting as the properties of $\mathcal{Z}_1$
are, their physical interpretation and application require care
and must be discussed in the context of the underlying model from
which $\mathcal{Z}_1$ was derived. In particular, a free Bose
field with a gapless spectrum is a delicate object which can
become unstable under infinitesimal perturbations (this should be
contrasted from a free Fermi field, which has a robust stability).
One instance of this instability is the response to an applied
magnetic field, $H_{\alpha}$; there must be a coupling which
imposes a precession of $\phi_{\alpha}$ about the direction of
magnetic field, and in these conditions the action is unstable to
arbitrarily large fluctuations in $\phi_{\alpha}$ in the
directions orthogonal to the field. For the initial spin glass
context in which $\mathcal{Z}_1$ was studied, the `quantum
critical' state described by (\ref{e1}) was found to be unstable
to the onset of spin glass order at low temperatures \cite{gps}.

In the remainder of this paper we will review another context in
which $\mathcal{Z}_1$ has appeared: the theory of quantum
impurities in insulators and superconductors with low energy
quantum spin fluctuations. Here, as we will see below, it is
essential to include a quartic $\phi_{\alpha}^4$ term for proper
computations of the $\hat{S}_{\alpha}$ correlations.

\section{Quantum antiferromagnets: $\phi^4$ field theory}
\label{phi4}

\begin{figure}
\centerline{\includegraphics[width=2.5in]{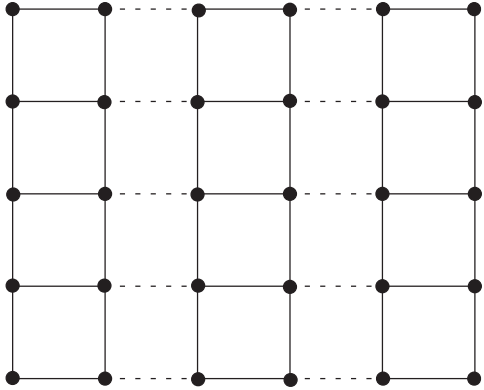}} \caption{The
coupled ladder Hamiltonian. Quantum spins resides on the filled
circles. They are coupled by two different antiferromagnetic
exchange constants, indicated by the full and dashed lines.}
\label{fig1}
\end{figure}

A concrete application of the `Bose Kondo' theory, which is now
reasonably well understood, is the problem of quantum impurities
in two-dimensional antiferromagnets. For definiteness, consider
the simple coupled ladder antiferromagnet, illustrated in
Fig~\ref{fig1}.
As the ratio of the exchange constants is varied, two distinct
types of ground states are obtained. For weakly-coupled ladders,
the ground state is a spin singlet and there is a gap to all
excitations; the ground state is adiabatically connected to the
state in which each spin is paired in a singlet with its partner
across the rung of the ladder. In contrast, when the inter- and
intra-ladder exchange constants are roughly equal, the model has
the structure of the square lattice antiferromagnet, and so has
antiferromagnetic N\'{e}el order in its ground state; in this case
spin rotation symmetry is broken, and the spin operators have an
average expectation value which has opposite signs on the two
sublattices. Given the distinct nature of these two ground states,
there must be a quantum phase transition between them. There is
now quite convincing evidence \cite{ladders} that there is one
second-order quantum critical point, and in its vicinity the spin
fluctuations are described by the $\phi^4$ field theory, written
here in $d$ spatial dimensions:
\begin{equation}
\widetilde{\mathcal{S}}_{\text{b}}[\phi_{\alpha}] = \int d^d x
d\tau \left[ \frac{1}{2} \left\{ (\partial_{\tau} \phi_{\alpha}
)^2 + c^2 (\nabla_x \phi_{\alpha})^2 + r \phi_{\alpha}^2 \right\}
+ \frac{u}{24} (\phi_{\alpha}^2)^2 \right]. \label{p4}
\end{equation}
The field $\phi_{\alpha}$ represents the staggered N\'{e}el order
parameter, and the tuning parameter $r$ moves the system from the
spin gap state at large $r$, to the N\'{e}el state at smaller $r$.
The spin-wave velocity is $c$, and $u$ is the quartic non-linear
coupling.

\begin{figure}
\centerline{\includegraphics[width=3.5in]{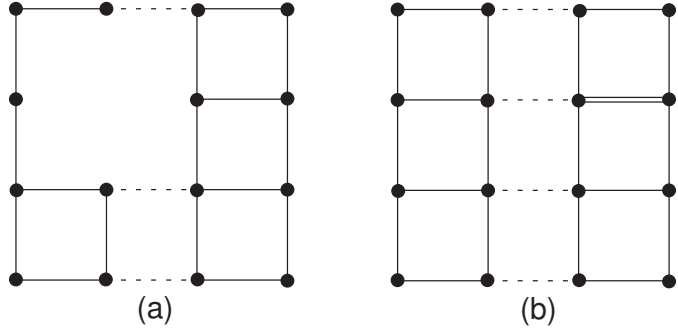}} \caption{Two
examples of quantum impurities in the coupled ladder
antiferromagnet, assumed to have spins with angular momentum
$S^{\prime}$ on the filled circles. ({\em a \/}) A vacancy, which
is characterized by $S=S^{\prime}$ in (\protect\ref{chiimp}).
({\em b\/}) A defect bond: the double line represents a large
ferromagnetic exchange, and this is characterized by
$S=2S^{\prime}$ in (\protect\ref{chiimp}).} \label{fig2}
\end{figure}

Now insert an arbitrary quantum impurity in the spin ladder
system: two examples are shown in Fig~\ref{fig2}.
In the spin gap state, the presence of such a impurity may
liberate one or more spins from their partners, and this leads to
a residual Curie spin susceptibility at a low temperature $T$:
\begin{equation}
\chi_{\text{imp}} = \frac{S(S+1)}{3 k_B T}~~~\mbox{;~~spin gap in
bulk antiferromagnet}. \label{chiimp}
\end{equation}
Here $S$ is an integer or half-odd-integer which characterizes the
impurity. A remarkable property \cite{sbv} of the low energy
dynamics of the quantum impurity is that {\em no other parameters}
are needed to described the spin dynamics in its vicinity,
provided the bulk antiferromagnet is not too far from its quantum
critical point. This result emerges from an analysis of the theory
coupling the impurity to the $\phi^4$ field theory:
\begin{eqnarray}
&&\mathcal{Z}_2 = \int \mathcal{D} \phi_{\alpha} (x, \tau)
\mathcal{ D} n_{\alpha} (\tau ) \delta \left( n_{\alpha}^2 - 1
\right) \exp\left(- \widetilde{\mathcal{S}}_{\text{b}}
[\phi_{\alpha}] - \widetilde{\mathcal{S}}_{\text{imp}} \right) \nonumber \\
&& \widetilde{\mathcal{S}}_{\text{imp}} =  \int d \tau \bigg[ i S
A_{\alpha} (n) \frac{d n_{\alpha}(\tau)}{d \tau}  - \lambda S
\phi_{\alpha} (x=0,\tau) n_{\alpha} (\tau) \bigg] , \label{simp2}
\end{eqnarray}
with $\widetilde{\mathcal{S}}_{\text{b}}[\phi_{\alpha}]$ given in
(\ref{p4}). Notice the similarity of $\mathcal{Z}_2$ to
$\mathcal{Z}_1$: at $r=0$, $u=0$, we can integrate out all the
$\phi_{\alpha} (x \neq 0, \tau)$, and then $\mathcal{Z}_2$ reduces
to $\mathcal{Z}_1$ with $\mu = d-1$. However, it is crucial in the
proper theory of $\mathcal{Z}_2$ that the non-linearity $u$ be
treated at an equal footing with $\lambda$; there is a non-trivial
`interference' between $u$ and $\lambda$, and the interaction $u$
significantly modifies the magnetic environment coupling to the
impurity. It is not permissible to treat the environment as a
Gaussian quantum noise, and focus only on its Kondo-like coupling
to the impurity.

A systematic renormalization group based analysis of
$\mathcal{Z}_2$ was carried out\cite{sbv} in an expansion in
$(3-d)$. The couplings $\lambda^2$ and $u$ both approach fixed
point values of order $(3-d)$, and this is the reason the coupling
between the bulk and impurity spin fluctuations becomes universal,
as claimed above. At the critical point, the spin correlations
decay as
\begin{equation}
\langle \hat{S}_{\alpha} (x,\tau) \hat{S}_{\alpha} (x,0) \rangle
\sim \frac{1}{|\tau|^{\eta^{\prime}}}~~~~~;~~~~~x \approx 0,
\label{e2}
\end{equation}
close to the impurity. The exponent $\eta^{\prime} \neq
2-\mu=3-d$, as would be implied by (\ref{e1}), because of the
non-zero fixed point value of $u$. Well away from the impurity,
the results are as in the absence of the impurity with
\begin{equation}
\langle \hat{S}_{\alpha} (x,\tau) \hat{S}_{\alpha} (x,0) \rangle
\sim \frac{1}{|\tau|^{d-1+\eta}}~~~~~;~~~~~x \rightarrow \infty,
\label{e3}
\end{equation}
where $\eta$ is the well-known anomalous dimensions of the
$\phi^4$ field theory in $d+1$ spacetime dimensions. The value of
$\eta^{\prime}$, and numerous other physical properties of the
impurity on both sides of the bulk quantum critical point, were
computed in Ref.~\cite{sbv} to second order in an expansion in
$(3-d)$. Numerical studies \cite{troyer,sandvik} have investigated
some of these properties. Related theoretical results were
obtained recently \cite{castro} in magnetically ordered states in
the presence of spin anisotropy.

\section{Quantum antiferromagnets: Non-linear sigma model}
\label{nls}

An alternative approach to the impurity dynamics discussed in
Section~\ref{phi4} is provided by a different representation of
the bulk spin fluctuations. It is well known that in low
dimensions the $\phi^4$ field theory can be represented by the
non-linear sigma model: the fluctuations of the amplitude,
$\phi_{\alpha}^2$, become irrelevant, and we need only focus on
the angular fluctuations of the N\'{e}el order parameter. These
are represented by a unit-length field $N_{\alpha} (x, \tau)$.
Such a `non-linear sigma model' representation provides an
expansion in powers of $(d-1)$ for the bulk critical properties.

In the quantum impurity problems of interest here, such a
fixed-length representation offers some benefits. One is that it
allows systematic computation of some properties in the
`renormalized classical' regime directly in spatial dimension
$d=2$. However, more importantly, in the fixed-length formulation
the universal nature of the couplings between the bulk and
impurity spin fluctuations can be accounted for at the outset.
Indeed, it was argued\cite{sv} that in the scaling limit of the
fixed-length theory, the quantum impurity behaves as if it is in
the $\lambda \rightarrow \infty$ limit, and hence the impurity
spin orientation align along the direction of the bulk spin order;
in other words, we have $n_{\alpha} (\tau) = N_{\alpha} (x=0,
\tau)$. With these arguments, we can rewrite the model described
by the partition function $\mathcal{Z}_2$ as
\begin{eqnarray}
&&\mathcal{Z}_3 = \int \mathcal{D} N_{\alpha} (x, \tau)
\delta\left(N_{\alpha}^2 - 1\right)
\exp\left(-\mathcal{S}_{\text{b}}
[N_{\alpha}] - \overline{\mathcal{S}}_{\text{imp}} \right) \nonumber \\
&& \overline{\mathcal{S}}_{\text{imp}} =  \int d \tau \bigg[ i S
A_{\alpha} (n) \frac{d n_{\alpha}(\tau)}{d \tau} \bigg]~~~
\mbox{with
$n_{\alpha} (\tau) = N_{\alpha} (x=0,\tau)$} \nonumber \\
&& \mathcal{S}_{\text{b}}[N_{\alpha}] = \frac{1}{2cg} \int d^d x
d \tau \left[ \left( \partial_{\tau} N_{\alpha} \right)^2 + c^2
\left( \nabla_x N_{\alpha} \right)^2 \right], \label{simp3}
\end{eqnarray}
where now $g$ is the coupling constant that tunes the bulk
antiferromagnet across the quantum critical point. Notice that
there is no other coupling constant, and hence the universal
nature of the coupling between the bulk and impurity is explicit.
A systematic $(d-1)$ expansion of $\mathcal{Z}_3$ was
performed\cite{sv}, and all results were found to be consistent
with those reviewed in Section~\ref{phi4}. In particular, a
$(d-1)$ expansion was presented for the exponent $\eta^{\prime}$
in (\ref{e2}), associated with computation of a `boundary'
renormalization constant for the field $N_{\alpha} (x=0, \tau)$.
Related results were also obtained in Ref.~\cite{sushkov2}.

\section{Electron spectral function in Kondo insulators}
\label{kondo}

Kondo insulators are another class of physically interesting
systems displaying a magnetic transition. In Kondo lattice models
with a commensurate density of conduction electrons, increasing
the Kondo exchange can drive the ground state from an ordered
N\'{e}el state to a paramagnetic insulator in which the conduction
electrons and local moments are strongly hybridized. The low
energy magnetic fluctuations near such a critical point are also
believed to be described by the $\phi^4$ field theory (\ref{p4}).

Now, let us consider the photoemission spectrum of a conduction
electron in such an insulator in the vicinity of the magnetic
transition \cite{assaad}. Away from the critical point, there will
be a sharp quasiparticle/hole pole with a finite residue, and the
position of this pole will disperse in the Brillouin zone. Focus
on the spectral function at the minimum of this
dispersion\cite{sushkov,stv}, where the momentum dependence is
quadratic. It was argued \cite{stv} that near this minimum, and in
the vicinity of the magnetic ordering transition where the action
(\ref{p4}) applies, we can safely neglect the quadratic dispersion
of the hole. We are therefore left with the problem of a {\em
static} hole interacting with the magnetic environment described
by (\ref{p4}). This problem is clearly analogous to the X-ray edge
problem, where a static hole interacts with a Fermi liquid.

This ``Bose X-ray edge'' problem is described by the partition
function
\begin{eqnarray}
&&\mathcal{Z}_4 = \int \mathcal{D} \phi_{\alpha} (x, \tau)
\mathcal{ D} \psi_a (\tau ) \mathcal{ D} \psi^{\dagger}_a (\tau )
\exp\left(- \widetilde{\mathcal{S}}_{\text{b}}
[\phi_{\alpha}] - \mathcal{S}_{\psi} \right) \nonumber \\
&& \mathcal{S}_{\psi} =  \int d \tau \bigg[ \psi_a^{\dagger}
\left( \frac{\partial}{\partial \tau} + \varepsilon_0
\right)\psi_a - \frac{\lambda}{2} \phi_{\alpha}(x=0, \tau)
\psi_a^{\dagger} \sigma^{\alpha}_{ab} \psi_b \bigg] ,
\label{simp4}
\end{eqnarray}
with $\widetilde{\mathcal{S}}_{\text{b}}[\phi_{\alpha}]$ given in
(\ref{p4}). The $\psi_a$ are Grassman variables representing the
hole with $a,b=\uparrow,\downarrow$, and $\sigma^{\alpha}$ are the
Pauli matrices. A complication has been ignored in our
presentation here of $\mathcal{Z}_4$ here: as the field
$\phi_{\alpha}$ carrier spin fluctuations at a finite momentum, it
actually couples fermionic excitations at two different points in
the Brillouin zone. We have not included this effect here because
keeping track of the fermionic momentum label does not modify the
critical properties\cite{stv}.

With the hole present, the quantum theory $\mathcal{Z}_4$ is, in
fact, identical to $\mathcal{Z}_2$ (with $S=1/2$): we have simply
realized the quantum spin by a single hole. Consequently, the
renormalization group equations for the $\lambda$ coupling in
(\ref{simp4}) are identical for those for $\lambda$ in
(\ref{simp2}). However, the present formulation allows
determination of a new renormalization constant associated with
the insertion of a hole. This constant measures the overlap of the
system wavefunctions with and without the hole; in contrast, in
(\ref{simp2}) the spin is always present and so this physics is
inaccessible.

Specifically, in the paramagnetic phase, away from the critical
point, the single hole Green's function $G$, has a quasiparticle
pole given by
\begin{equation}
G(\omega) = \frac{Z}{\omega-\varepsilon_0} \label{g}
\end{equation}
where $Z$ is the quasiparticle residue, and $\varepsilon_{0}$ has
absorbed a renormalization from the coupling of the hole to
$\phi_{\alpha}$. As we approach the critical point, there is an
orthogonality catastrophe and $Z \rightarrow 0$. Instead, at the
critical point we have\cite{stv,sushkov,iss}
\begin{equation}
G(\omega) \sim \frac{1}{(\omega - \varepsilon_0 )^{1 - \eta_f}}.
\label{g1}
\end{equation}
The exponent $\eta_f$ is distinct from $\eta^{\prime}$, and its
determination requires a separate renormalization group analysis.
The value of $\eta_f$ has been obtained in a two-loop expansion
\cite{iss} in powers of $(3-d)$, and by a numerical simulation
\cite{stv}.

\begin{acknowledgments}
I thank Chiranjeeb Buragohain, Antoine Georges, Olivier Parcollet,
Matthias Troyer, Matthias Vojta, and Jinwu Ye for fruitful
collaborations on the topics reviewed here. This research was
supported by US NSF Grant DMR 0098226.
\end{acknowledgments}

\end{document}